\documentclass{IEEEcsmag}

\usepackage{amsmath,amsfonts} 
\usepackage{upmath,tabularx,booktabs,bm,subfigure}
\newcommand{\kai}[1]{\textcolor{black}{{#1}}}
\jvol{XX}
\jnum{XX}
\paper{8}
\jmonth{May/June}
\jname{IT Professional}
\pubyear{2019}

\newcommand{\m}{{dEFEND}}
\begin{document}

\editor{Editor: VS Subrahmanian, vs@dartmouth.edu}

\title{Detecting Fake News with Weak Social Supervision}





\author{

\IEEEauthorblockN{
Kai Shu\IEEEauthorrefmark{1},
Ahmed Hassan Awadallah\IEEEauthorrefmark{2},
Susan Dumais\IEEEauthorrefmark{2}, and
Huan Liu\IEEEauthorrefmark{1}
}
\IEEEauthorblockA{
\IEEEauthorrefmark{1}Arizona State University, \{kai.shu, huan.liu\}@asu.edu\\
\IEEEauthorrefmark{2}Microsoft Research, \{hassanam, sdumais\}@microsoft.com}}


\begin{abstract}
\kai{Limited labeled data is becoming one of the largest bottlenecks for supervised learning systems. This is especially the case for many real-world tasks where large scale labeled examples are either too expensive to acquire or unavailable due to privacy or data access constraints. 
Weak supervision has shown to be effective in mitigating the scarcity of labeled data by leveraging weak labels or injecting constraints from heuristic rules and/or extrinsic knowledge sources.} Social media has little labeled data but possesses unique characteristics that make it suitable for generating weak supervision, resulting in a new type of weak supervision, i.e., \textit{weak social supervision}. In this article, we illustrate how various aspects of social media can be used as weak social supervision. 
\kai{Specifically, we use the recent research on fake news detection as the use case, where social engagements are abundant but annotated examples are scarce, to show that weak social supervision is effective when facing the labeled data scarcity problem. This article opens the door to learning with weak social supervision for similar emerging tasks when labeled data is limited.}

\end{abstract}

\maketitle

\begin{IEEEkeywords}
\kai{social media, weak supervision, social networking}
\end{IEEEkeywords}


\kai{
\chapterinitial{Social media} has become an important means of large-scale information sharing and communication in all occupations, including marketing, journalism, public relations, and more. Due to the increased usage and convenience of social media, more people seek out and receive  timely  news  information  online.  For  example,  the  Pew  Research  Center  announced  that approximately  68\%  of  US  adults  get  news  from  social  media  in  2018,  while  only  49\%  reported seeing  news  on  social  media  in  2012.  However,  social  media  also  proliferates  a  plethora  of  misinformation  and  disinformation,  including fake  news,  i.e.,  news  stories  with  intentionally  false information~\cite{shu2019detecting}. For example, during the 2016 U.S. election, the top most frequently-discussed false stories generated 8,711,000 shares, reactions, and comments on Facebook, larger than the total of 7,367,000 top most-discussed true stories.
Detecting fake news on social media is critical to avoid people to consume false information and cultivate a healthy and trustworthy news ecosystem.}

However, detecting fake news on social media poses several unique challenges~\cite{shu2019detecting}. First, the \textit{data challenge} has been a major roadblock for researchers in their attempts to develop effective defensive means against disinformation and fake news. This is because the content of fake news and disinformation is rather diverse in terms of topics, styles and media platforms; and fake news attempts to distort the truth with diverse linguistic styles while simultaneously mocking true news. Thus, obtaining labeled fake news data is non-scalable and data-specific embedding methods are not sufficient for fake news detection with little labeled data. Second, the \textit{evolving challenge} of disinformation and fake news makes it non-trivial to exploit the rich auxiliary information signals directly. Fake news is usually related to newly emerging, time-critical events, which may not have been properly verified by existing knowledge bases (KB) due to the lack of corroborating evidence or claims. \kai{Moreover, detecting fake news at an early stage requires the prediction models to utilize minimal information from user engagements because extensive user engagements indicate more users are already affected by fake news.}

\kai{Recently, learning with weak supervision has been of great interest the research community to mitigate the data scarcity problem for various tasks
.  Social media data has unique properties that make it suitable for deriving weak supervision. }
First, social media data is \textit{big}. We have limited data for each individual. However, the social property of social media data links individuals’ data together, which provides a new type of big data.
Second, social media data is \textit{linked}. The availability of social relations determines that social media data is inherently linked, meaning it is not independent and identically distributed.
Third, social media data is \textit{noisy}. Users in social media can be both passive content consumers and active content producers, causing the quality of \kai{user-generated} content to vary. Social networks are also noisy with the existence of malicious users such as spammers and bots.
Therefore, social media data provides a new type of weak supervision, \textit{weak social supervision}, which has great potentials to advance a wide range of applications including fake news detection.



\kai{
In this article, we propose a new type of weak supervision from multi-faceted social media data, i.e., weak social supervision, and illustrate how to effectively derive and exploit the weak supervision for learning with little labeled data. 
We discuss three major perspectives of the social media data to derive weak social supervision for fake news detection: users, posts, and networks. Further, we introduce recent work on exploiting weak social supervision for \textit{effective} and \textit{explainable} fake news detection. First, we illustrate how we can model the relationships among publishers, news pieces, and social media users with user-based and network-based weak social supervision to detect fake news effectively.  Second, we show how to leverage post-based weak social supervision for discovering explainable comments while detecting fake news. Finally, we discuss several open issues and provide future directions of learning with weak social supervision.}

\section{\kai{Learning with Weak Supervision}}
Learning with weak supervision is an important and newly emerging research area, and there are different ways of defining and approaching the problem.
One definition of weak supervision is leveraging higher-level and/or noisier input from subject matter experts (SMEs)
. The supervision from SMEs are represented in the form of weak label distributions, which mainly come from the following sources: 1) \textit{inexact supervision}: a higher-level and coarse-grained supervision; 2) \textit{inaccurate supervision}: a low-quality and noisy supervision; and 3) \textit{existing resources}: using existing resources to provide supervision. Another \kai{definition} categorizes weak supervision into inexact supervision, inaccurate supervision, and incomplete supervision~\cite{zhou2017brief}. \kai{The incomplete supervision means that a subset of training data are given with labels, which essentially includes active learning and semi-supervised learning techniques. }
Weak supervision can be formed in deterministic (e.g., in the form of \textit{weak labels}) and non-deterministic (e.g., in the form of \textit{constraints}) ways.

\subsubsection{Incorporating Weak Labels}
Learning with noisy (inaccurate) labels has been of great interest to the research community for various tasks. Some of the existing works attempt to rectify the weak labels by incorporating a loss correction mechanism. 
Patrini \textit{et al.}~\cite{patrini2017making} utilize the loss correction mechanism to estimate a label corruption matrix without making use of clean labels. Other works consider the scenario where a small set of clean labels are available. For example, \kai{Veit \textit{et al.} use human-verified labels and train a label cleaning network in a multi-label classification setting~\cite{veit2017learning}}. 
In some cases, weak supervision is obtained with inexact labels such as coarse-grained labels. For example, object detectors can be trained with images collected from the web using their associated tags as weak supervision instead of locally-annotated data sets.

\subsubsection{Injecting Constraints}
Directly learning with weak labels may \kai{suffer from} the noisy label problem. Instead, representing weak supervision as constraints can avoid noisy labels and encode domain knowledge into the learning process of prediction function. The constraints can be injected over the output space and/or the input representation space. For example, Stewart \textit{et al.}~\cite{stewart2017label} model prior physics knowledge on the outputs to penalize ``structures'' that are not consistent with the prior knowledge. For relation extraction tasks, label-free distant supervision can be achieved via encoding entity representations under transition law from knowledge bases (KB). This type of weak supervision, i.e., injecting constraints, is often based on prior knowledge from domain experts, which are \textit{jointly} optimized with the primary learning objective of prediction tasks.

\section{Learning with Weak Social Supervision}
In the previous section, we introduced the definitions and techniques for learning with weak supervision. In this section, we further formally define the problem of learning with weak social supervision, introduce how to derive weak social supervision and exploit it for fake news detection.

\subsection{\kai{From Weak Supervision to Weak Social Supervision}}
With the rise of social media, the web has become a vibrant and lively realm where billions of individuals all around the globe interact, share, post and conduct numerous daily activities. Social media enables us to be connected and interact with anyone, anywhere and anytime, which allows us to observe human behaviors in an unprecedented  scale with \kai{a} new lens. However, significantly different from traditional data, social media data is big, incomplete, noisy, unstructured, with abundant social relations. This new type of data contains rich \textit{social interactions} that can provide \kai{additional} signals for obtaining weak supervision. Next, we formally define the problem of learning with weak social supervision.


A training example consists of two parts: a feature vector (or \textit{instance}) describing the event/object, and a \textit{label} indicating the ground-truth. Let $\mathcal{D}=\{x_i, y_i\}_{i=1}^{n}$ denote a set of $n$ examples, with $\mathcal{X}=\{x_i\}_{i=1}^{n}$ denoting the instances and $\mathcal{Y}=\{y_i\}_{i=1}^{n}$ the corresponding labels. In addition, there is a large set of unlabeled examples. Usually the size of the labeled set $n$ is much smaller than the unlabeled set due to labeling costs or privacy concerns.  

For the widely available unlabeled samples, we generate weak social supervision by generating weak labels or incorporating constraints based on social media data. For weak labels, we aim to learn a labeling function $g: \tilde{\mathcal{X}}\rightarrow \tilde{\mathcal{Y}}$, where $\tilde{\mathcal{X}}=\{\tilde{x}_j\}_{j=1}^{N}$ denotes the set of $N$ unlabeled messages to which the labeling function is applied and $\tilde{\mathcal{Y}}=\{\tilde{y}_j\}_{j=1}^N$ as the resulting set of weak labels. This weakly labeled data is then denoted by $\tilde{\mathcal{D}}=\{\tilde{x}_j, \tilde{y}_j\}_{j=1}^{N}$ and often $n<<N$. For formulating constraints, we aim to model prior knowledge from social signals on the representation learning of examples 
with a constraint function $h:\mathcal{X}\times\mathcal{Y}\rightarrow\mathbb{R}$, to penalize structures that are not consistent with our prior knowledge.
Note that $g$ can also be applied to $\tilde{\mathcal{X}}$ to regularize the representation learning. In spite of the different forms we model weak social supervision, we are actually aiming to estimate a label distribution $p(\tilde{y}|\tilde{x})$ from weak social supervision. We give the following problem formulation of learning with weak social supervision.

\begin{center}
\fbox{\parbox[c]{.9\linewidth}{\textbf{Learning with Weak Social Supervision:}
Given little data with ground truth labels $\mathcal{D}$ and a label distribution $p(\tilde{y}|\tilde{x})$ derived from weak social supervision,  learn a prediction function $f: \mathcal{X}\rightarrow \mathcal{Y}$ which generalizes well onto unseen samples.
}}
\end{center}

\subsection{\kai{Deriving Weak Social Supervision}}
Next, we illustrate how to derive weak social supervision for fake news detection. Generally, there are three major aspects of the social media context: users, generated posts, and networks. 

\subsubsection{User-based:}  \kai{Fake news pieces are likely to be created and spread by non-human accounts, such as social bots or cyborgs~\cite{kumar2016disinformation}}. Thus, capturing users' profiles and characteristics as weak social supervision can provide useful information for fake news detection. \kai{User behaviors can indicate their characteristics~\cite{subrahmanian2017predicting} who have interactions with the news on social media. These signals can be categorized in different levels: \textit{individual-level} and \textit{group-level}~\cite{shu2019detecting}. Individual-level signals are extracted to infer the credibility and reliability for each user using various aspects of user demographics, such as registration age, number of followers/followees, number of tweets the user has authored, etc. Group-level user signals capture overall characteristics of groups of users related to the news. The injected constraints  of weak supervision is that the spreaders of fake news and real news may form different communities with unique characteristics that can be depicted by group-level signals. }

\subsubsection{Post-based:} 
\kai{Users who are involved in news dissemination process express their opinions, emotions via posts/comments, which provide helpful signals related to the veracity of news claims~\cite{shu2019defend}. 
Recent research looks into \textit{stance}, \textit{emotion}, and \textit{credibility} to improve the performance of fake news detection~\cite{shu2019detecting}. First, stances (or viewpoints) indicate the users' opinions towards the news, such as supporting, opposing, etc. Typically, fake news can provoke \kai{tremendously} controversial views among social media users, in which denying and questioning stances are found to play a crucial role in signaling claims as being fake~\cite{shu2019detecting}. 
Second, fake news publishers often \kai{aim} to spread disinformation extensively and draw wide public attention. Longstanding social science studies demonstrate that the news which evokes high-arousal, or activating (awe, anger or anxiety) emotions is more viral on social media~\cite{berger2012makes}.  
Third, post credibility aims to infer the veracity of news pieces from the credibility of the posts on social media. The injected constraint of weak supervision  is that the credibility of the news is highly related to the credibility degree of its relevant social media posts. }

\subsubsection{Network-based:} 
\kai{Users form different networks on social media in terms of interests, topics, and relations. Fake news dissemination processes tend to form an echo chamber cycle, highlighting the value of extracting network-based weak social supervision to represent these types of network patterns for fake news detection~\cite{shu2019detecting}. Different types of networks can be constructed such as \textit{friendship networks}, \textit{diffusion networks}, \textit{interaction networks}, etc. 
First, friendship network plays an important role in fake news diffusion. The fact that users are likely to form echo chambers~\cite{quattrociocchi2016echo}, strengthens our need to model user social representations and to explore its added value for a fake news study. 
Second, the news diffusion process involves abundant temporal user engagements on social media.  Fake news may have \kai{a sudden increase in} the number of posts and then remain constant beyond a short time whereas, in the case of real news, the increase of the number of posts are more steady~\cite{shu2019detecting}. 
In addition, an important problem along temporal diffusion is the early detection of fake news with limited amount of user engagements. 
Third, interaction networks describe the relationships among different entities such as publishers, news pieces, and users.  For example, the user-news interactions are often modeled by considering the relationships between user representations and the news veracity values. 
Intuitively, users with low credibilities are more likely to spread fake news, while users with high credibility scores are less likely to spread fake news~\cite{shu2019beyond}.  
}

\subsection{Exploiting Weak Social Supervision}
Earlier, we illustrate different aspects that we can \kai{derive} weak social supervision from. \kai{It is worth mentioning that the extracted weak social supervision can involve single or multiple aspects of the information related to users, content, and networks.} In this section, we discuss learning with weak social supervision for fake news detection in different settings including \textit{effective fake news detection} and  \textit{explainable fake news detection}. Specifically first, we illustrate how we can model the user-based and network-based weak social supervision to detect fake news effectively. Second, we show how to leverage post-based weak social supervision for discovering explainable comments for detecting fake news. 

\subsubsection{Effective Fake News Detection}
\begin{figure}[tb]

        \center{\includegraphics[width=0.45\textwidth]
        {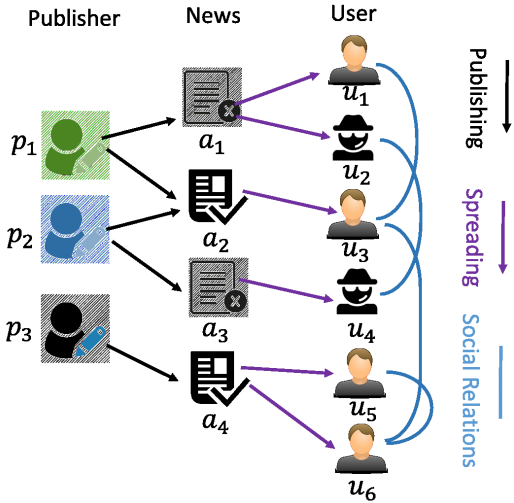}}
        \caption{\label{fig:interaction} \kai{An illustration of the relationships among publishers, news pieces, and users, which can be modeled as weak social supervision to detect fake news~\cite{shu2019beyond}.}}

\end{figure}

We aim to leverage weak social supervision as an auxiliary information to perform fake news detection effectively. As an example, we demonstrate how we can utilize interaction networks by modeling the entities and their relationships to detect fake news (see Figure~\ref{fig:interaction}). Interaction networks describe the relationships among different entities such as publishers, news pieces, and users. Given the interaction networks the goal is to embed the different types of entities into the same latent space, by modeling the interactions among them. We can leverage the resultant feature representations of news to perform fake news detection, and we term this framework \underline{Tri}-relationship for \underline{F}ake \underline{N}ews detection (TriFN)~\cite{shu2019beyond}. 


Social science research has demonstrated the following observations which \kai{provide motivations to derive rules of} weak social supervision: \kai{\textit{people tend to form relationships with like-minded friends, rather than with users who have opposing preferences and interests~\cite{quattrociocchi2016echo}.}} Thus, connected users are more likely to share similar latent interests in news pieces. for publishing relationship, we exploit the following weak social supervision: \textit{publishers with a high degree of political bias are more likely to publish fake news.} Moreover, for the spreading relation, we have: \textit{users with low credibilities are more likely to spread fake news, while users with high credibility scores are less likely to spread fake news}. . We utilize nonnegative matrix factorization (NMF) to learn the news representations by encoding the weak social supervision. \kai{Specifically, the label distribution $p(\tilde{y}|\tilde{x})$ is estimated by \textit{injecting constraints} into the heterogeneous network embedding framework for learning the news representations: (1) for publishing relationship, we enforce that the news representation should be good at predicting the partisan bias of its publisher; (2) for the spreading relationship, we constrain that the news presentation and user representation are close to each other if the news is fake and the user is less-credible, and vice versa.}

\begin{figure}[tb]
        \center{\includegraphics[width=0.45\textwidth]
        {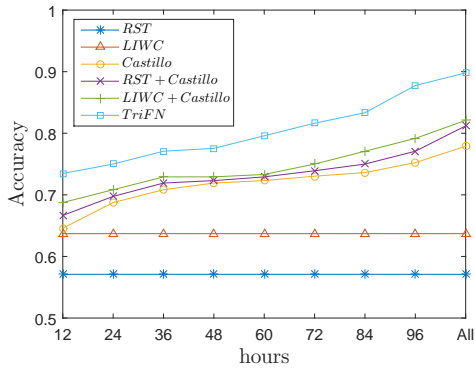}}
        \caption{\label{fig:wsdm} The performance of fake news detection with different amount of weak social supervision. No weak social supervision is incorporated in RST, LIWC, while Castillo only encodes weak social supervision. TriFN, which utilizes both labeled data and weak social supervision, can achieve the best performance.}
\end{figure}

\textbf{Empirical Results} To illustrate whether the weak social supervision in TriFN can help \kai{to detect} fake news effectively, we show some empirical comparison results in the public benchmark Politifact dataset from FakeNewsNet (\texttt{github.com/KaiDMML/FakeNewsNet}) as in Figure~\ref{fig:wsdm}, \kai{which consists of 120 true news and 120 fake news pieces, with 91 publishers and 23,865 users}. \kai{We compare TriFN with baselines that 1) only extract features from news contents, such as RST~\cite{rubin2015towards}, LIWC~\cite{pennebaker2015development}; 2) only construct features from social supervision, such as Castillo~\cite{castillo2011information}; and 3) consider both news content and social supervision, such as RST+Castillo, LIWC+Castillo. In particular, (1) RST~\cite{rubin2015towards} stands for Rhetorical Structure Theory, which builds a tree structure to represent rhetorical relations among the words in the text; (2) LIWC~\cite{pennebaker2015development} stands for Linguistic Inquiry and Word Count, which is widely used to extract the lexicons falling into psycholinguistic categories. It's based on a large sets of words that represent psycholinguistic processes, summary categories, and part-of-speech categories; (3) Castillo~\cite{castillo2011information} extract various kinds of features from those users who have shared a news item on social media. The features are extracted from user profiles and friendship network. We also include the credibility score of users  as an additional social context feature.} 

We can see that the proposed TriFN can achieve around 0.75 accuracy even with a limited amount of weak social supervision (within 12 hours after the news is published), and has as high as 0.87 accuracy. In addition, with the help of weak social supervision from publisher-bias and user-credibility, the detection performance is better than those without utilizing weak social supervision. Moreover, we can see within a certain range, more weak social supervision leads to the larger performance increase, which shows the  benefit of using weak social supervision.

\subsubsection{Explainable Fake News Detection}
In recent years, computational detection of fake news has been producing some promising early results. However, there is a critical missing piece of the study, the explainability of such detection, i.e., why a particular piece of news is  detected as fake. Here, we introduce how we can derive explanation factors from weak social supervision.

\begin{figure}[tb]
   \centering
   \includegraphics[scale=0.34]{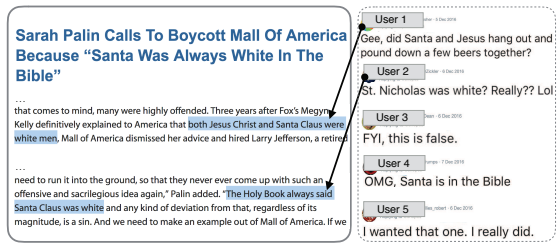}
   \caption{\kai{A piece of fake news with related user comments on social media. Some explainable comments are directly corresponding to the sentences in news contents.}}
   \label{fig:case_study_news}
   \vspace{-1.2em}
\end{figure}

We observe that not all sentences in news contents are fake, and in fact, many sentences are true but only for supporting \kai{the false} claim sentences. Thus, news sentences may not be equally important in determining and explaining whether a piece of news is fake or not. 
Similarly, user comments may contain relevant information about the important aspects that explain why a piece of news is fake, while they may also be less informative and noisy. \kai{For example, in Figure~\ref{fig:case_study_news}, we can see users discuss different aspects of the news in comments such as ``\texttt{St. Nicholas was white? Really??Lol},'' which directly responds to the claims in the news content ``\texttt{The Holy Book always said Santa Claus was white}.'' }

Therefore, we use the following weak social supervision: \textit{the user comments that are related to the content of original news pieces are helpful to detect fake news and explain prediction results}. \kai{The label distribution $p(\tilde{y}|\tilde{x})$ is also estimated by injecting constraints such that: \kai{semantically} related news sentences and user comments are attended to predict and explain fake news. }
Thus, we aim to select some news sentences and user comments that can explain why a piece of news is fake. As they provide a good explanation, they should also be helpful in detecting fake news. This suggests us to design attention mechanisms to give high weights of representations of news sentences and comments that are beneficial to fake news detection. Specifically, we first use Bidirectional LSTM with attention to learn sentence and comment representations, and then utilize a sentence-comment co-attention neural network framework called dEFEND to exploit both news content and user comments to jointly capture explainable factors.

\begin{figure}[htb!]
        \center{\includegraphics[width=0.49\textwidth]
        {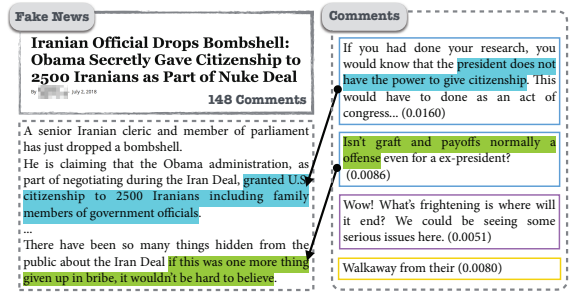}}
        \caption{\label{fig:kdd2} \kai{The case study of leveraging weak social supervision for explanation.} }
\end{figure}

\textbf{Empirical Results} 
\begin{figure}[tb]
        \center{\includegraphics[width=0.45\textwidth]
        {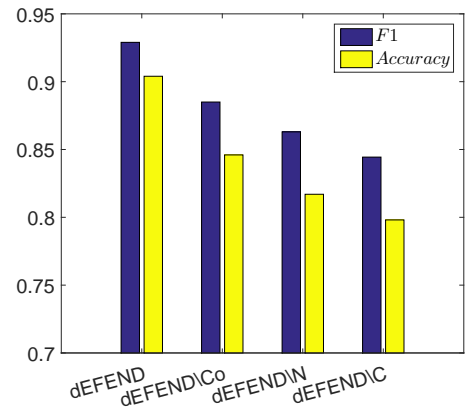}}
        \caption{Assessing the effects of news contents and weak social supervision based on user comments.}
        \label{fig:kdd1}
\end{figure}
\kai{We show the empirical results on Politifact platform from FakeNewsNet as in Figure~\ref{fig:kdd1}, which consists of 145 true news and 270 fake news pieces, with 89,999 comments from 68,523 users. The labels are manually assigned by journalist experts from the fact-checking websites such as \texttt{PolitiFact.com}, and the social interactions such as users and their comments are collected from Twitter.}  We can see {\m} achieves very high performances in terms of accuracy ($\sim$ 0.9) and F1 ($\sim$ 0.92). We compare {\m} with three variants: 1) {\m}\textbackslash C not considering information from user comments; 2) {\m}\textbackslash N is not considering information from news contents; and 3) {\m}\textbackslash Co eliminates the sentence-comment co-attention. We observe that when we eliminate news content component, user comment component, or the co-attention for news contents and user comments, the performances are reduced. It indicates capturing the semantic relations between the weak social supervision from user comments and news contents are important. \kai{The evaluation of explainability includes the perspectives of news sentences and user comments. First, the Mean Average Precision (MAP) is adopted as the metric to evaluate how explainable are news sentences. The results indicate that dEFEND can achieve better MAP scores than baselines such as HAN~\cite{yang2016hierarchical}. Second, we use Amazon Mechanical Turk to perform human
evaluation on ranking the explainable comments, and Normalized Discounted Cumulative Gain
(NDCG) as the metric. We observe dEFEND can achieve better NDCG performance to capture explainable comments than baselines. 
}
\kai{
Moreover, we also illustrate the case study of using weak social supervision as an explanation in Figure~\ref{fig:kdd2}. We can see that: {\m} can rank more explainable comments higher than non-explainable comments. For example, comment ``\texttt{...president does not have the power to give citizenship...}'' is ranked at the top, which can  explain exactly why the sentence ``\texttt{granted U.S. citizenship to 2500 Iranians including family members of government officials}'' in the news content is fake;
In addition, we can give higher weights to explainable comments than those interfering and unrelated comments, which can help select more related comments to help detect fake news. For example, unrelated comment ``\texttt{Walkaway from their...}'' has an attention weight $0.0080$, which is less than an explainable comment ``\texttt{Isn't graft and payoffs normally a offense}'' with an attention weight $0.0086$.
}

\section{Open Issues and Future Research}
In this section, we present some open issues in weak social supervision and future research directions. 

\subsection{Weak Social Supervision for Fake News Detection}
Most of the current methods are trying to exploit weak social supervision as \textit{constraints} to help fake news detection. We can also exploit generating \textit{weak labels} from the aforementioned social signals (user-based, post-based, and network-based) as labeling functions for early fake news detection. 
\kai{Some representative labeling rules for extracting weak labels are described as follows~\cite{shu2020mining}
: 1) \textit{Credibility}-based: \textit{users with low credibilities are more likely to spread fake news, while users with high credibilities are less likely to spread fake news}; 2) \textit{Bias}-based: \textit{publishers with more partisan bias are more likely to publish fake news than mainstream publishers with less bias}; 3) \textit{Sentiment}-based: \textit{news with more conflicting viewpoints in the related media posts tends to be fake than those with less conflicting viewpoints}. Empirical results show that by exploiting these multi-sources of weak social supervision can significantly improve the detection performance with limited labeled data of fake news.}
The advantage of leveraging weak social supervision for early fake news detection is that we can jointly learn the feature representations from little labeled data and weakly labeled data, and when predicting unseen news pieces, we can perform prediction with few/no social signals, which perfectly satisfy the requirement of early detection. In addition, in the extreme case when no labeled data \kai{is} available, we can utilize weak social supervision for unsupervised fake news detection. One idea is to extract users' opinions on the news by exploiting the auxiliary information of the users' engagements from posts on social media, and aggregate their opinions in a well-designed unsupervised way to generate our estimation results. 

\subsection{Techniques for Learning Weak Social Supervision} We expect along the direction of learning with weak social supervision, more research will emerge in the near future. First, leveraging weak social supervision for computation social science research is promising. Since computational social science research usually relies on relatively limited offline survey data, weak social supervision can serve as a powerful online resources to understand and study social computing problems. Second, existing approaches utilize single or combine multiple sources of weak social supervision, while to what extent and aspect the weak social supervision helps is fairly important to explore. Third, the capacity of ground-truth labels and weak social supervision and the relative importance between the sources are essentials to develop learning methodology in practical scenarios. Moreover, the weak supervision rules may have complementary information since they capture social signals from different perspectives. An interesting future direction is to explore multi sources of weak social supervision in a principled way to model the mutual benefits through data programming.

\section{Conclusion}
In many machine learning applications, labeled data is scarce and obtaining more labels is expensive. Motivated by the promising early results of exploiting weak supervision learning, we propose a new type of weak supervision, i.e., \textit{weak social supervision}. We specifically focus on the use case of detecting fake news on social media. Specifically, We demonstrate that weak social supervision provides a new representation to describe social information uniquely available where a better warning is sought, which has promising results and great potentials toward detecting fake news, including challenging settings of effective fake news detection and explainable fake news detection.  We also further discuss promising future directions in fake news detection research and expand the field of learning with weak social supervision to other applications.


\bibliographystyle{IEEEtran}
\bibliography{ref,ref_weak}

\begin{IEEEbiography}{Kai Shu}{\,}is a PhD student and research assistant at Data Mining and Machine Learning (DMML) Lab at Arizona State University. His research interests include artificial intelligence, social computing, data mining. Contact him at kai.shu@asu.edu.
\end{IEEEbiography}

\begin{IEEEbiography}{Ahmed Hassan Awadallah}{\,}is a principle research manager at Microsoft Research (MSR). He leads the language and information technology team at MSR, focusing on creating language understanding and user modeling technologies to enable intelligent experiences in multiple products. Contact him at hassanam@microsoft.com.
\end{IEEEbiography}

\begin{IEEEbiography}{Susan Dumais}{\,}is a the technical fellow and Managing Director, Microsoft Research New England, New York City and Montreal. Her research interests are in algorithms and interfaces for improved information retrieval, as well as general issues in human-computer interaction. Contact her at sdumais@microsoft.com.
\end{IEEEbiography}

\begin{IEEEbiography}{Huan Liu}{\,}is a professor of Computer Science and Engineering at Arizona State University. His research interests are in data mining, machine learning, social computing, and artificial intelligence, investigating problems that arise in real-world applications with high-dimensional data of disparate forms. Contact him at huan.liu@asu.edu.
\end{IEEEbiography}

\end{document}